\documentstyle[prl,aps,epsf,multicol]{revtex}
\begin{document}
\topmargin -15mm
\title{Ground and Metastable States in $\gamma$-Ce from Correlated Band Theory}

\author{A. B. Shick$^a$, W. E. Pickett$^{a}$ and A. I. Liechtenstein$^{b}$}

\address{$^a$Department of Physics, University of California, Davis, CA 95616 \\
$^b$Department of Physics, University of Nijmegen, the Netherlands} 

\maketitle

\begin{abstract}
{Multiple energy minima of the LDA+U energy functional are obtained for $\gamma$-Ce
when it is implemented in a full potential, rotationally invariant scheme including
spin-orbit coupling, and different starting local configurations are chosen.
The lowest energy solution leads to a fully spin polarized 4f state and the
lattice constant of $\gamma$-Ce.  The higher energy local minima 
(additional self-consistent solutions) are shown to
be strongly indicative of crystal electric field and multiplet excitations.
}
\end{abstract}

\begin{multicols}{2}

The isostructural $\alpha$ - $\gamma$ phase transition in 
Ce, and the character of the phases separately, have been a subject of
continious theoretical efforts for decades. Experimentally, the low temperature
fcc $\alpha$ phase of Ce transforms to the fcc $\gamma$ phase at the
high temperatures with a large increase of the volume. When pressure is
applied, the crystal collapses back into the $\alpha$ phase in a first-order
phase transition. The room temperature $\gamma-\alpha$ phase transition occurs
at $\approx$ 7 kbar pressure with the decrease of the volume of $\approx$ 15\%.

Several theoretical models have been suggested and they
are described in some detail
in  Ref. \cite{Svane2}. The difference between the models arises from the different
treatments for the Ce 4f electrons. The promotional model \cite{Coqblin} 
assumed depopulation of the 4f level upon compression, but does not agree with the
results of various experiments which indicate little change in the 
4f occupation.  Band structure calculations also indicate that promotion
is too high in energy to drive the transition, and the 4f occupation
remains near unity.\cite{Pickett}
The Mott transition model \cite{Johansson}
assumes the transformation of localized to band-like 4f state with the
decrease of volume and is similar to the Mott-Hubbard metal-insulator transition.
First principles calculations using the self interaction
correction (SIC) to the LDA \cite{Svane} produce total energy
minima for localized ($\gamma$) and band ($\alpha$)
states with formally the same SIC total energy variational
functional, and are in accord with the Mott transition model. However, the use of
atomic sphere approximation (ASA), which sphericalizes the potential and density
within large atomic spheres, limits the ability of these calculations to
produce quantitative total energy description of the system with 
highly non-spherically symmetric
4f electron charge/spin densities.
 
The other viable theoretical model is the Kondo volume collapse (KVC)
model \cite{Allen}. The essence of the KVC is an assumption of localized f states in
the both $\alpha$ and $\gamma$ phases of Ce. The $\alpha$ phase consists of 
a mixed valence 4f state while the $\gamma$ phase has almost integer 4f occupancy.
The phase transition is due to the entropy contribution of
the localized 4f state. The relation between KVC and Mott transition model was
analyzed recently in Ref. \cite{Svane2} and quantitative arguments in favor of
KVC model were provided.

The local (spin) density approximation (LDA) has had tremendous success in 
the quantitative description of a wide varietly of solids,\cite{Jones} but the rare
earths are extreme cases where the LDA description is inadequate.  The 
``correlated band theory'' (LDA+U) approach has had much success in the 
treatment of correlated magnetic insulators where LDA results are incorrect.
In this paper we explore not the $\gamma-\alpha$ transition {\it per se},
but the $\gamma$ phase itself, by an application of 
the LDA+U  approach that includes the new features (1) a full potential
method is used, which enables the novel features we uncover, and (2) the LDA+U
method is applied to a metal where the interaction of the local orbitals with
the conduction states is a central part of the physics.  For the first time,
multiple LDA+U energy minima (ground plus metastable states) are obtained within 
a specified broken symmetry, and we discuss how the various states are related
to 4f excitations in the $\gamma$ phase.

The LDA+U method \cite{Anisimov} is based on an energy which is a functional of the spin densities
\{$\rho^s, s=\pm 1$\}, and the occupation 
matrices \{$n^s_{m,m^{\prime}}$\} of the 4f orbitals
labelled by their azimuthal projections $m \equiv m_{\ell}$.  There is every
reason to expect that, within the allowed values of $\rho^s$ 
and $n^s$, there can 
be local energy minima as well as the absolute minimum, which represents
the ground state energy of the system. (Such local minima are rarely found
within LDA except in magnetic systems.)  
The formal meaning of these local minima,
as well as the formal underpinnings of the LDA+U approach itself, remain to be
settled, but -- like the Kohn-Sham eigenvalues (the band structure) which have
little formal meaning but immense practical importance -- these minima will be shown to
bear a close relationship to local excitations of the 4f shell.

Our results are based upon the full-potential linearized augmented 
plane wave method (LAPW)\cite{Singh} as the basis for
total energy calculations with the rotationally invariant LDA+U
functional \cite{flapwu}, with spin-orbit
coupling (SOC) is included self-consistently \cite{flapwso}. 
Literature values \cite{Anisimov2} are used for the on-site repulsion $U \; = \; 6.1 $ eV
and exchange $J \; = \; 0.7$ eV
(Slater integrals, $F_0 =$ 6.10 eV, $F_2 =$ 8.34 eV, $F_4 =$ 5.57 eV, $F_6 =$ 4.12 eV ).
Ferromagnetic spin alignment is assumed (allowed, not imposed)
since local moments are present in paramagnetic
high temperature $\gamma$ phase of Ce \cite{Gschneidner}.

The key to finding the various solutions (local minima) is to 
start from atomic densities and different `guesses' for the 4f occupation matrices $n^s$
and then obtain self-consistently both spin/charge densities and 
occupation matrices.
As an intuitive guess for 4f occupations we use the fully 
spin-polarized state (Tr $n^{\uparrow}$=1, Tr $n^{\downarrow}$=0) 
and choose various
orbital characters for $n^{\uparrow}_{m,m^{\prime}}$. 

{\it Ground State.}
The calculated equilibrium lattice constant (cf. Table I)
is very close to the experimental value of $\gamma$-Ce at zero
pressure and room temperature \cite{Savrasov}.
The calculated bulk modulus is about 25\% larger than the experimental value,
similar to other calculations that have treated the localized character of the
4f state in $\gamma$-Ce.

For the $\gamma$ phase volume (lattice constant a=9.76 a.u.) 
we find the lowest solution
to be fully spin polarized with the f occupation of 1.04. 
This state is primarily $m_l$=-2 
($m_j=-\frac{3}{2})$, with some mixing in of the spin-majority $m_l=2$ state.
The LDA+U yields significant enhancement for the absolute values of both spin and orbital
magnetic moments (anti-aligned in accord with the 3rd Hund's rule) in comparison with
LDA results.

The electronic density of states for the $\gamma$-Ce ground state is shown in Fig. 1 in comparison with
the result of relativistic (with SOC) LDA calculations. The f-majority peak at the
bottom of valence band ($\approx$ 2.5 eV below the Fermi level) indicates the position of the localized
4f state, in quantitative agreement with the resonant photoemission measurements \cite{Weschke}.
The difference between LDA+U and LDA is due to
the localization of the 4f state, which removes both majority and minority 4f states from the vicinity of the
Fermi level.  
Since the 4f state is not well separated from the valence band (cf. Fig. 1) we
conclude that in spite of its localized character the 4f 
state in $\gamma$-Ce cannot be
treated as core-like.

{\it Metastable States.}
As mentioned above, several
self-consistent solutions corresponding to different (local) minima
of the LDA+U energy functional are possible.
Different minima must be searched for by exploring various regions of the
underlying space.  Our initial starting points consisted of fully spin
polarized 4f$^1$ with different orbital characters \{$m_l$\}.
Our initial and final (self-consistent) 
majority spin occupation matrix elements are shown in 
Table II.
In the LDA+U calculations the spin polarization enters as an effective
magnetic field providing the spin quantization axis
and the direction of the quantization axis is chosen along
z([001]).

The LDA+U minima can be considered as mean-field-like 
projections (in the Fock space) of
many-body wavefunctions on the single-particle angular basis set. The
interpretation is much simpler for an f$^1$ ion than it would
be for a multi-electron ion because the $\{m_l,m_s\}$ and $\{J,J_z\}$ representations
are unitarily related. In the simple crystal-electric-field (CEF) model for the paramagnetic
Ce$^{3+}$ ion \cite{Fulde} the ground state 
is formed by one of the  $\Gamma_7$ and $\Gamma_8^{1,2}$ doublets from $J=5/2$ multiplet \cite{CEF}. 
In order to compare the CEF model \cite{Fulde} with the LDA+U results, we apply to the LDA+U solutions
the unitary transformation
from $\{m_l,m_s\}$ representation to $\{J,J_z\}$ representation
with $J=5/2,7/2$.
The states from Table III are then classified as follows:

\noindent {\it Solution (1) - Ground State}: It has 96\% $m_l=-2$,
with 3\% $m_l=2$ mixed in. The transformation to $\{J,J_z\}$ representation
yields  69\% of $|5/2,-3/2>$ and 1\% of $|5/2,5/2>$ states from $J=5/2$ multiplet,
plus 27\% of $|7/2,-3/2>$ and 3\% of $|7/2,5/2>$ states from
$J=7/2$.
We conclude that the {\it Solution 1} consists of
70\% states from the  $J=5/2$ multiplet which are the linear combination
of the CEF levels $\Gamma_7$ and $\Gamma_8^1$, and 30\% states
from the $J=7/2$ multiplet.
 
\noindent {\it Solution (2)}: Roughly equal amounts of $m_l=-3$ and $m_l=+1$) 
in this solution indicates coupled $m_j=-5/2$ and $m_j=3/2$ states.
It consists of 55\% states from $J=5/2$ multiplet (combinations
of  the CEF levels $\Gamma_7$ and $\Gamma_8^1$) and 
45\% of states from $J=7/2$.

\noindent {\it Solution (3)}, involving 
admixture of $m_l=-1$ with 25\% $m_l=+3$,
which transforms to an admixture of 42\% of $|5/2,-1/2>$ ($|\Gamma_8^2>$) state with the 58\% of
$|7/2,7/2>$.

\noindent {\it Solution (4)}, pure $m_l=0$, corresponds to the
combination of 42\% of $|5/2,1/2>$ ($|\Gamma_8^2>$) state with the 58\% of
$|7/2,1/2>$, 
and lies 232 meV above the ground state.

These correlated band structure 
calculations are producing metastable solutions that
contain the CEF states from the conventionally presumed lowest multiplet for  Ce$^{3+}$.
In addition, there is a considerable fraction of the states from
the first excited $J_{7/2}$ multiplet that must be understood.

There are two important reasons why our solutions do not correspond 
directly to CEF levels as normally considered. First, the CEF picture assumes
ideal cubic symmetry, whereas a 4f state in $\gamma$-Ce is actually surrounded
by twelve atoms whose own moments are oriented randomly, hence breaking cubic symmetry.
In addition the non-cubic nature of the 4f density perturbes the conduction electron
density, which leads to a non-cubic local field. Secondly, in our calculations we have artificially ordered
the spins of the magnetic ions and allowed orbital moments, which reduces out site symmetry to
tetragonal.

The comparison between LDA+U calculated splitting scheme with the results 
of inelastic neutron scattering experiments  \cite{Murani} is shown in Fig. 2.
It is clearly seen that {\it Solutions 1-3}
lie well within the range of low-energy
excitations peak and reflect the mixed CEF and spin-orbit excitations. The energy position
of {\it Solution 4} (232 meV) agrees quantitatively
with the experimental inelastic peak at ~260 meV. This peak is usually interpreted 
as the spin-orbit excitation $^2F_{5/2} \; \rightarrow \; ^2F_{7/2}$ \cite{Murani}.
Our calculations suggest that this interpretation is oversimplified since the states from both
lower $^2F_{5/2}$ and first excited $^2F_{7/2}$ multiplets are mixed  in
the ground and excited states of $\gamma$-Ce.
The spectrum from present LDA+U calculations
are in  better agreement with the experiment \cite{Murani} than previously reported
results of SIC calculations \cite{Temmerman} (86-100 meV and 130 meV).

In order to classify the LDA+U solutions
we transform them from the compex to the cubic spherical harmonics.
{\it Solution (1)} involves admixture
of $xyz$ and $(x^2-y^2)z$ states; {\it Solutions (2) and (3)}
consist of mixed $(x^2-z^2)y$ and $(y^2-z^2)x$ states;
and {\it Solution (4)} has $(5z^2-3r^2)z$ symmetry.
All the LDA+U solutions have the tetragonal symmetry assumed in the
calculations. The localized states without SOC
are formed by CEF states in $\{L,L_z\}$ representation and for $L=3$
these are $\Gamma_2,\Gamma_4,\Gamma_5$ states \cite{Yosida}.
In fully spin-polarized case the spin-orbit coupling $\xi ({\vec{l} \cdot \vec{s}})$ is reduced to
$\xi ({\hat{l}_z \hat{s}_z})$ and the eigenvalues of localized ``CEF+SOC"
model Hamiltonian are easily obtained. 
The ``CEF+SOC" coupling scheme is presented in Fig. 3 and shows how 
the LDA+U solutions result from CEF eigenstates coupled
by spin-orbit coupling. From the total energy differences
between LDA+U solutions we then derive the CEF splitting parameters $\Delta_{2,5} \; = \; \Gamma_2 - \Gamma_5$ = 90 meV
and $\Delta_{4,2} \; = \; \Gamma_4 - \Gamma_2$ = 107 meV and SOC parameter $\xi$ of 66 meV (SOC splitting = $(7/2) \xi$ =
231 meV).

The scheme in Fig. 3 allows more eigenstates than the calculated local minima
of the LDA+U total energy functional. In each case the occupation matrices $n_{m_l,m_l'}$ for
these additional states have the non-zero elements in the same $\{m_l,m_l'\}$ sub-space as one of
the metastable states.
As a result the minimization procedure yields the lowest total 
energy state in given $\{m_l,m_l'\}$ sub-space and the higher 
states do not appear among variational LDA+U solutions.

The physical origin of the CEF splitting can be understood to be
due to the anisotropy of the mixing interaction between
conduction band and localized f-states \cite{Levy}.
 We obtain the CEF splitting for $\gamma$-Ce
of the order of 100 meV  comparable with the SOC splitting
of 231 meV. In both ground and excited states the CEF
levels are then mixed by SOC yielding low-energy excitations
of the order of 20-50 meV and high-energy excitations about
230 meV.

The results of LDA+U calculations allow us to conclude
that the interaction between conduction band and localized
f-states plays an important role for the both ground and 
excited states in $\gamma-$Ce. As a result, the Ce f-states
are {\it not core-like} and should be treated in terms of
quantum impurity (Anderson) model \cite{PWA} rather than
as localized states of Ce$^{3+}$ ion in the cubic
CEF. The present LDA+U calculations correspond to the numerical solution
of the lattice Anderson model in static mean-field approximation with the assumed ferromagnetic order.
This simplification does not allow us to describe  quantitatively the paramagnetic (disordered local
moment) high-temperature state of $\gamma-$Ce observed experimentally.
The treatment of this paramagnetic state 
requires the use of dynamic
mean-field  parametrization for the self-energy\cite{Katsnelson}.

Recently, Solovyev {\it et al.} \cite{Solovyev} proposed a correction
to the LDA+U total energy functional \cite{ldau} when SOC is included
and non-collinear magnetic configurations are considered.
This correction accounts for additional contributions to the exchange energy
due to non-zero spin-off-diagonal elements of the occupation matrix 
\{$n^{s,-s}_{m,m^{\prime}}$\}. We performed  LDA+U calculations with this
spin-off-diagonal correction \cite{Solovyev} and found that the spin-off-diagonal occupations are
very small and have minute effects on the values of spin and orbital magnetic
moments, and  the total energies (less than 2 meV) for ground and
metastable states (cf. Table II) in $\gamma-$Ce.    


To summarize, we have obtained the ground and three metastable states 
from  relativistic (with spin-orbit coupling)
spin-polarized full-potential LDA+U  calculations for fcc Ce.
The ground state has equilibrium lattice constant
of $\gamma$-Ce.
Our calculations reproduce 
the localized character of 4f states in $\gamma$-Ce, which 
however cannot be treated as a part
of atomic core.  Analysis of various LDA+U solutions allows us to
make a comparison between correlated band theory results and
CEF model. The excitation energies calculated from the total energy differences between
LDA+U solutions are in reasonable agreement with the experimental data.

We are grateful to D. L. Cox for helpful comments.
This research was supported by
National Science Foundation Grant DMR-9802076, and by NSF Grant PHY-94-07194
while A.I.L and W.E.P. were in residence at the Institute of Theoretical Physics,
UCSB.



{\narrowtext
\begin{figure}
\label{ldad}
\epsfxsize=7cm
\epsffile{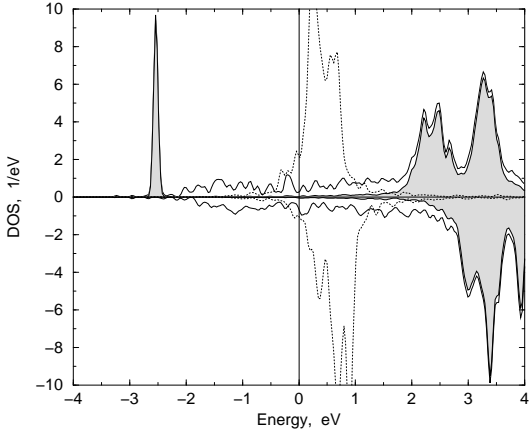}
\caption{Spin-resolved total and partial 4f
densities of states for ordered $\gamma$-Ce.  Majority are plotted upwards, minority downwards.
The 4f partial DOS from LDA is dotted.  The full line gives the LDA+U
DOS, with the 4f contribution filled in.  The 4f DOS peaks near the
Fermi level in both spins are well removed from the Fermi level by
LDA+U.}
\end{figure}
}

{\narrowtext
\begin{figure}
\label{ldad}
\epsfxsize=7cm
\epsffile{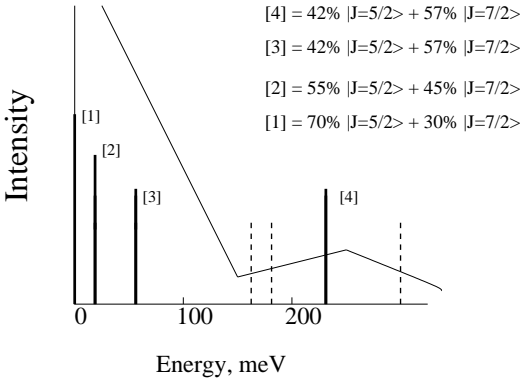}
\caption{Schematic sketch of the experimental inelastic neutron scattering  
intensity in comparison with the energy positions
of LDA+U calculated ground and excited states [1]-[4].
Additional inferred localized excitations are shown (dotted lines).} 
\end{figure}
}

{\narrowtext
\begin{figure}
\label{ldad}
\epsfxsize=7cm
\epsffile{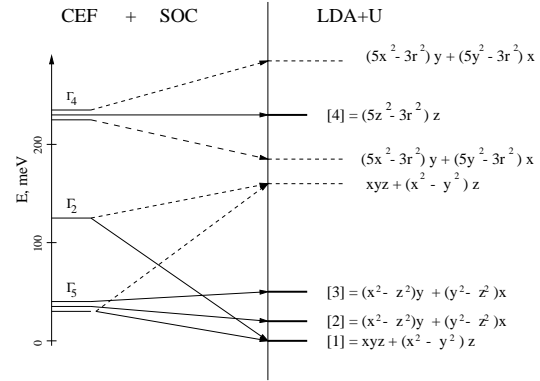}
\caption{A schematic sketch for the relation of the CEF levels 
to the LDA+U calculated ground and excited states [1]-[4] (full lines).
Additional possible localized excitations are shown (dotted lines).} 
\end{figure}
}



\narrowtext
\begin{table}
\caption
{The equilibrium lattice constant ($a, a.u.$) and bulk moduli ($B, kbar$) as a result
of LDA+U calculations. 
}
\begin{tabular}{cccccccccccc}
           &LAPW                    &LMTO-ASA \cite{Svane} & \multicolumn{2}{c}{LMTO-ASA \cite{Johansson2}}& Exp. \cite{Svane} \\
           & LDA+U                  & SIC                  & f-core LDA & f-core GGA                    &           \\
$a$        &9.83                    & 9.58                 & 9.69       & 10.02                         & 9.76 \\
$B$        &296                     & 310                  & 312        & 288                           & 210, 244 \\
\end{tabular} 
\label{tab1}
\end{table}

\begin{table}
\caption{
The elements of the occupation matrix $n_{m,m'}$ (spin - majority) (initial assignment $n^0$ and self-consistent $n^{scf}$);
spin ($2<S_z>$), orbital ($<L_z>$) moments and the total energy increase with
respect to the ground state ($\Delta E, meV$) for different LDA+U self-consistent solutions 
in the order of increasing total energy.
The only those elements of occupation matrix which are 
bigger than 0.01 are shown. The  spin-minority occupation matrix is almost zero since there is a
complete spin-polarization of the 4f-shell. 
}
\begin{tabular}{ccccccccccc}
\multicolumn{3}{c}{$n^{0}_{m,m'}$}& \multicolumn{3}{c}{$n^{scf}_{m,m'}$}& $2<S_z>$ & $<L_z>$ & $\Delta E$ \\
\hline
\multicolumn{6}{c}{Solution 1} \\
\hline
$m,m'$ & -2 & 2 & $m,m'$ & -2    & 2            & 1.18     &-1.87    & 0       \\
-2     &  1& 0 & -2  & 0.960 & 0.143 \\
 2     &  0& 0 &  2  & 0.143 & 0.030  \\
\hline
\multicolumn{6}{c}{Solution 2} \\
\hline
$m,m'$ & -3 & 1 & $m,m'$ & -3    & 1            & 1.18     &-0.75    & 19        \\
-3     &  1& 0 & -3  & 0.438 &-0.484 \\
 1     &  0& 0 &  1  &-0.484 & 0.552  \\  
\hline
\multicolumn{6}{c}{Solution 3} \\
\hline
$m,m'$ & -1 & 3 & $m,m'$ & -1    & 3            & 1.18     & 0.001   & 52        \\
-1     &  1& 0 & -1  & 0.737 &-0.423 \\
 3     &  0& 0 &  3  &-0.423 & 0.253  \\  
\hline
\multicolumn{6}{c}{Solution 4} \\
\hline
$m,m'$ &  0 &   & $m,m'$ &  0    &              & 1.14     &-0.004   & 232       \\
 0     &  1 &   &  0     & 0.979 &\\
\end{tabular} 
\end{table}

\end{multicols}

\end{document}